\documentclass[12pt,aps,prl,superscriptaddress]{revtex4}
\usepackage{amsmath,amssymb,graphicx}

\newcommand{\Omicron}{\mathrm{O}}
\newcommand{\mathe}{\mathrm{e}}
\newcommand{\tmem}[1]{{\em #1\/}}
\newcommand{\tmop}[1]{\ensuremath{\operatorname{#1}}}
\newcommand{\tmtextbf}[1]{{\bfseries{#1}}}
\newcommand{\tmtextit}[1]{{\itshape{#1}}}
\newcommand{\tmtexttt}[1]{{\ttfamily{#1}}}

\begin{document}

\title{Precision check on triviality of $\phi^4$ theory by a new simulation
method}
\author{Ulli Wolff}
\affiliation{Institut f\"ur Physik, Humboldt Universit\"at,
      Newtonstr. 15, 12489 Berlin, Germany}

\begin{abstract}
  We report precise simulations of $\phi^4$ theory in the Ising limit. A
  recent technique to stochastically evaluate the all-order strong coupling
  expansion is combined with exact identities in the closely related Aizenman
  random current representation. In this way estimates of the
  renormalized coupling close to the continuum limit
  become possible with unprecedented precision and yet low CPU cost. As a sample application
  we present results for the unbroken phase of the Ising model in dimensions
  3, 4 and 5 and investigate the question of triviality by studying a finite
  size scaling continuum limit.
\end{abstract}
\begin{flushright} HU-EP-09/04 \end{flushright}
\begin{flushright} SFB/CCP-09-20 \end{flushright}
\maketitle

The $\phi^4$ theory of a real scalar field is the starting point
of many textbooks on quantum field theory. It also plays a phenomenological
r{\^o}le as an extremely simplified model of the Higgs sector of the standard model.
In \cite{Luscher:1987ay}, building on \cite{Brezin:1976bp}, it was discussed that the theory is trivial in
dimension four, meaning that no interaction can exist in the continuum limit.
For applications this means, that the effective theory {\em with} an unremovable cutoff
in place has only a limited energy domain of validity.
This important property is rigorously known to hold above four dimensions 
\cite{Aizenman:1981du}, \cite{Froehlich1982281}, but in $D=4$ we have to rely so-far
on numerical checks. In this letter we report on the discovery of a new numerical
strategy and algorithm, which enhances the precision of such checks
by orders of magnitude as the continuum limit is approached. Also novel is
the use of a finite volume renormalization scheme in this context. Due to both improvements we
can progress much deeper
into the universal scaling region than in previous computations like \cite{Montvay:1988uh}.
We find a `borderline' agreement with standard perturbation theory for the cutoff
dependence of the interaction strength which calls for more study.

We here consider the Ising limit of $\phi^4$
on a $D$-dimensional hypercubic lattice of extent
$L$ and lattice spacing $a$ in all directions. We mostly use lattice units $a
= 1$ from here on, but occasionally re-introduce $a$ to emphasize cutoff
dependencies. Physical information is extracted
via $n$-point correlation functions
\begin{equation}
  \left\langle s (x_1) s (x_2) \cdots s \left( x_n \right) \right\rangle =
  \frac{Z (x_1, x_2, \ldots, x_n)}{Z (\emptyset)} \label{scorr}
\end{equation}
with
\begin{equation}
  Z (x_1, x_2, \ldots, x_n) = 2^{- V} \sum_s \mathe^{\beta \sum_{l = \langle x
  y \rangle} s (x) s (y)} s (x_1) s (x_2) \cdots s \left( x_n \right)
\end{equation}
and the volume $V = L^D$. We sum over all Ising configurations $s (x) = \pm 1$
and $Z (\emptyset)$ is the proper partition function with no field insertions.
On our finite lattice $Z (.)$ is analytic for all values of $\beta$. We
parametrize the strong coupling expansion in $\beta$ by summing in addition to
$s$ over an integer link field $k (l) = 0, 1, \ldots, \infty$
\begin{equation}
  Z (x_1, x_2, \ldots, x_n) = 2^{- V} \sum_{s, k} w [k] \prod_{l = \langle x y
  \rangle} [s (x) s (y)]^{k (l)} s (x_1) s (x_2) \cdots s \left( x_n \right)
\end{equation}
with the multiple Poisson weight
\begin{equation}
  w [k] = \prod_l \frac{\beta^{k (l)}}{k (l) !} .
\end{equation}
For each $k$ the $s$ sum can now be performed and leaves behind a constraint
for $k$,
\begin{equation}
  Z (x_1, x_2, \ldots, x_n) = \sum_k w [k] \delta_{Q [k], X} .
\end{equation}
The Kronecker $\delta$ enforces the coincidence of two sets. The source set $Q
[k]$ consists of the sites surrounded by an {\tmem{odd}} total number of $k
(l)$,
\begin{equation}
  Q [k] = \left\{ x |  \sum_{l, \partial l \ni x} k (l) = 1 (\tmop{mod} 2)
  \right\} \label{Qkdef} .
\end{equation}
The insertion set $X$ coincides with $(x_1, x_2, \ldots, x_n)$ if they are
mutually different, but is more generally given by
\begin{equation}
  X = \left\{ x |  \sum_{i = 1}^n \delta_{x, x_i} = 1 (\tmop{mod} 2) \right\}
  \label{Xdef} .
\end{equation}

Michael Aizenman {\cite{Aizenman:1981du}}, {\cite{Aizenman:1982ze}} has used
the above representation of the Ising model to obtain rigorous correlation
inequalities. He calls $\{k (l)\}$ random currents and the sets $Q = X$
defects or external sources. Among his results the following is of interest
here as it can be turned into an efficient numerical algorithm. From
proposition 5.1 in {\cite{Aizenman:1982ze}} the identity
\begin{equation}
  Z_c (x_1, x_2, x_3, x_4) = - 2 \sum_{k, k'} w [k] w [k'] \delta_{Q [k],
  X_{12}} \delta_{Q [k'], X_{34}} \mathcal{X}(x_1, x_3 ; k + k') \label{Aizen}
\end{equation}
with the connected part of $Z_c$ of $Z$
\begin{eqnarray}
  Z_c (x_1, x_2, x_3, x_4) & = & Z (x_1, x_2, x_3, x_4) - Z (x_1, x_2) Z (x_3,
  x_4) \nonumber\\
  & - & Z (x_1, x_3) Z (x_2, x_4) - Z (x_1, x_4) Z (x_2, x_3)  \label{con}
\end{eqnarray}
follows. The sets $X_{12}$ and $X_{34}$ are formed as in (\ref{Xdef}) but from
only two points each. The cluster incidence function $\mathcal{X} \in \{0,
1\}$ is one if $x_1, x_3$ are in the same bond percolation cluster built by
bonds that are active on links where $k (l) + k' (l) > 0$ holds in the doubled
random current system. Note that the pairs $\{x_1, x_2 \}$ and $\{x_3, x_4 \}$
are connected automatically for $k, k'$ that contribute.

In the Monte Carlo community is has recently been found
{\cite{prokofev2001wacci}}, {\cite{Deng:2007jq}}, {\cite{Wolff:2008km}} that
it is both possible and advantageous to simulate the untruncated strong
coupling expansion instead of the original path integral (or sum) over fields.
In a simple variant of the worm algorithm {\cite{prokofev2001wacci}} one
simulates the ensemble corresponding to the partition function
\begin{equation}
  \mathcal{Z}= \sum_{u, v, k} w [k] \delta_{Q [k], X_{u v}} \label{Zens}
\end{equation}
with a corresponding definition of expectation values of observables 
$\left\langle \left\langle\mathcal{O}[u, v, k]\right\rangle \right\rangle$.
In the sum $u, v$ run over all lattice sites.

To simulate this ensemble our elementary update step is as follows:
\begin{itemize}
  \item pick at random one of the $2 D$ links emanating from $u$ and call it
  $l = \langle u \tilde{u} \rangle$
  
  \item assign a new value $\tilde{k} (l)$ to this link with probability
  $p_{\tilde{k}} = \exp (- \beta) \beta^{\tilde{k}} / \tilde{k} !$
  
  \item if $\tilde{k} - k$ is odd, move $u \rightarrow \tilde{u}$, otherwise
  leave $u$ unchanged
\end{itemize}
If we alternate these steps with similar ones for $v$ we have a correct
algorithm for (\ref{Zens}). Ergodicity may be shown by steps deforming an
arbitrary configuration to the trivial one. This local heatbath has proved to
be slightly superior to Metropolis proposals with $k (l) \rightarrow k (l) \pm
1$. It is not difficult to show that the Ising two-point function is now given
by the ratio of histograms
\begin{equation}
  \left\langle s (x) s (0) \right\rangle = \frac{\left\langle \left\langle
  \delta_{x, u - v} \right\rangle \right\rangle}{\left\langle \left\langle
  \delta_{u, v} \right\rangle \right\rangle} \label{twopoint}
\end{equation}
where we have used translation invariance. This implies in particular that the
susceptibility is given by
\begin{equation}
  \chi_2 = \sum_x \left\langle s (x) s (0) \right\rangle = [ \left\langle
  \left\langle \delta_{u, v} \right\rangle \right\rangle]^{- 1} .
\end{equation}
To now make use of (\ref{Aizen}) for the connected four point susceptibility
\begin{equation}
  \chi_4 = \sum_{x, y, z} \left\langle s (x) s (y) s (z) s (0) \right\rangle_c
\end{equation}
with subtractions as in (\ref{con}) all we have to do is simulate two
independent replica of (\ref{Zens}) and sum over all $x_i$ in (\ref{Aizen}) to
arrive at
\begin{equation}
  -_{} V \chi_4 = 2 \frac{\left\langle \left\langle \mathcal{X}(u, u', k + k')
  \right\rangle \right\rangle}{\left\langle \left\langle \delta_{u, v}
  \delta_{u', v'} \right\rangle \right\rangle} .
\end{equation}
In total we thus have derived
\begin{equation}
  - \frac{1}{V} \frac{\chi_4}{(\chi_2)^2} = 2 \left\langle \left\langle
  \mathcal{X}(u, u', k + k') \right\rangle \right\rangle .
\end{equation}
The right hand side is obviously bounded between 0 and 2. In particular the
lower bound corresponds to the Lebowitz inequality. Our estimator reflects
this property manifestly, the subtraction of disconnected parts has been
achieved analytically. We expect this to lead to a superior precision for
$\chi_4$ compared to conventional Monte Carlo procedures since they involve
substantial numerical cancellations here with the correspondingly enhanced
relative errors.

The previous expression is strongly reminiscent of a standard definition of a
dimensionless universal renormalized coupling constant in $\phi^4$ theory
including the Ising limit with its infinite bare coupling. It is given by
\begin{equation}
  g_R = - \frac{\chi_4}{(\chi_2)^2} m^D \label{gR}
\end{equation}
where $m$ is a the renormalized mass.

Often, for example in {\cite{Brezin:1976bp}}, {\cite{Luscher:1987ay}}, the
mass is defined in terms of the two point function in an infinite volume at
vanishing momentum. We substitute this by a definition using the two smallest
possible momenta in a periodic volume as in {\cite{Deng:2007jq}}. The two
point function (\ref{twopoint}) in momentum space may be measured by
\begin{equation}
  \tilde{G} (p) = \frac{\left\langle \left\langle \cos (p (u - v))
  \right\rangle \right\rangle}{\left\langle \left\langle \delta_{u, v}
  \right\rangle \right\rangle} .
\end{equation}
Our definition of a renormalized mass $m$ is
\begin{equation}
  \frac{m^2}{m^2 + \hat{p}^2_{\ast}} = \frac{\tilde{G} (p_{\ast})}{\tilde{G}
  (0)} = \left\langle \left\langle \cos (p_{\ast} (u - v)) \right\rangle
  \right\rangle
\end{equation}
where we use the smallest momentum
\begin{equation}
  p_{\ast} = (2 \pi / L, 0, 0, \ldots, 0), \hspace{1em} a \hat{p}_{\ast} = 2
  \sin (\pi a / L)
\end{equation}
and average over its $D$ possible directions. The rationale of the definition
(\ref{gR}) is that it is a dimensionless ratio with the same number of fields
in the numerator and denominator and hence it is expected to have a universal
continuum limit. As it vanishes for Gaussian theories it is a measure of
the interaction strength. We are thus led to the definition
\begin{equation}
  g = 2 \left\langle \left\langle \mathcal{X}(u, u', k + k') \right\rangle
  \right\rangle \times z^D, \hspace{1em} z = m L.
\end{equation}

Combining triviality with finite size scaling we investigate the proposition
that the continuum limit at fixed $z$ forces $g \searrow 0$. As for other
questions on non-perturbative ultraviolet renormalization
{\cite{Wolff:1985pj}}, {\cite{Luscher:1991wu}}, {\cite{Luscher:1992an}} we
find it advantageous to employ a finite volume renormalization scheme also
here. We shall perform a sequence of simulations of growing $L \equiv L / a$
where we tune $\beta$ such as to maintain a fixed value $z = 2$. The advantage
of keeping $L$ finite and not too large in physical length units $m^{- 1}$ is
that for the manageable values of $L / a$ we expect to be closer to the
universal continuum limit. If the theory is trivial we should find $g
\rightarrow 0$ as $L / a \rightarrow \infty$. This is expected
{\cite{Aizenman:1981du}}, {\cite{Aizenman:1982ze}}, {\cite{Froehlich1982281}}
for $D > 4$, and likely, although only at a logarithmic rate, for $D = 4.$

\begin{table}[htb]
  \begin{tabular}{|l|l|l|l|l|l|l|}
    \hline
    $D$ & $L / a$ & $\beta$ & $z$ & $\mathcal{X}$ & $\partial \mathcal{X}/
    \partial z$ & $\mathcal{X}(z = 2)$\\
    \hline
    4 & 8 & 0.148320 & 1.9981(27) & 0.39235(96) & -0.3200(14) & 0.39175(63)\\
    \hline
    4 & 10 & 0.148748 & 1.9949(26) & 0.37256(92) & -0.3193(14) & 0.37093(62)\\
    \hline
    4 & 12 & 0.148996 & 1.9992(26) & 0.35493(91) & -0.3165(15) & 0.35469(60)\\
    \hline
    4 & 16 & 0.149270 & 1.9988(25) & 0.33161(91) & -0.3129(16) & 0.33125(58)\\
    \hline
    4 & 22 & 0.149449 & 2.0085(24) & 0.30831(86) & -0.3030(16) & 0.31088(57)\\
    \hline
    4 & 32 & 0.149571 & 1.9956(24) & 0.29028(83) & -0.2993(20) & 0.28896(55)\\
    \hline
    3 & 8 & 0.217350 & 1.9946(36) & 0.59387(100) & -0.2929(13) & 0.59228(76)\\
    \hline
    3 & 10 & 0.218560 & 1.9942(37) & 0.58634(102) & -0.2950(13) &
    0.58463(76)\\
    \hline
    3 & 16 & 0.220153 & 2.0047(37) & 0.57240(109) & -0.3023(14) &
    0.57382(77)\\
    \hline
    3 & 32 & 0.221143 & 2.0032(39) & 0.56338(118) & -0.3076(17) &
    0.56435(76)\\
    \hline
    5 & 8 & 0.113052 & 2.0041(18) & 0.19126(59) & -0.2390(13) & 0.19223(45)\\
    \hline
    5 & 10 & 0.113336 & 1.9993(16) & 0.16037(53) & -0.2170(13) & 0.16022(41)\\
    \hline
    5 & 12 & 0.113503 & 1.9937(15) & 0.13884(47) & -0.1957(12) & 0.13760(38)\\
    \hline
    5 & 16 & 0.113674 & 1.9918(13) & 0.10944(39) & -0.1656(12) & 0.10809(33)\\
    \hline
  \end{tabular}
  \caption{Simulation results for $D = 3, 4, 5$.\label{tab}}
\end{table}

To get confidence in our algorithmic implementation we first reproduced the
results of cluster simulations in {\cite{Montvay:1988uh}} within errors. For
the $20^4$ lattice our error in $g$ for a comparable number of Flops is about
12 times smaller.

In the table we compile our data. Each line corresponds to $10^6$ iterations
with $V$ link updates in each of the two replica. The cost is dominated by the
runs $D = 4, L = 32$ and $D = 5, L = 16$ with 240 hours each on a single PC.
We refer to a code running under \tmtexttt{matlab} but importing random
numbers from the \tmtexttt{ranlux} generator {\cite{Luscher:1993dy}} in C
{\cite{ranlux}}. We have insisted on luxury level 2, but this part still
accounts for only 4 \% of the CPU time. The main code could clearly be
accelerated substantially. Derivatives of $\mathcal{X}$ and $z$ with respect
to $\ln \beta$ can be measured as connected correlations with $\sum_l k (l)$
and their quotient yields our estimate for $\partial \mathcal{X}/ \partial z$.
In the end it emerges as a certain nonlinear function of primarily measured
observables and it as well as all other errors is estimated by the tools
provided in {\cite{Wolff:2003sm}}. We take its measured value in each data set
as a fixed constant and then form, now with $\mathcal{X}$ and $z$ as functions
of primary quantities, the combination
\begin{equation}
  \mathcal{X}(z = 2) =\mathcal{X}+ (2 - z) \partial \mathcal{X}/ \partial z .
\end{equation}
A look at the table shows that this is a tiny but sometimes significant
correction. We can however safely neglect the error of $\partial \mathcal{X}/
\partial z$ and higher terms of the Taylor expansion. It came as a pleasant
surprise that even where no systematic correction is needed the statistical
fluctuations in this combination partially cancel and thus reduce the error.
This saves more than another factor two in run-time. The compensation is
actually plausible: when sampled graphs are `bigger' than average
$\mathcal{X}$ goes up but the mass goes down. In the table we note that
relative errors are practically independent of $L / a$, which means complete
absence of critical slowing down. Nevertheless the (short) autocorrelations do
have to be taken into account when errors are determined.

\begin{figure}[htb]
\begin{center}
  \includegraphics[width=0.8\columnwidth]{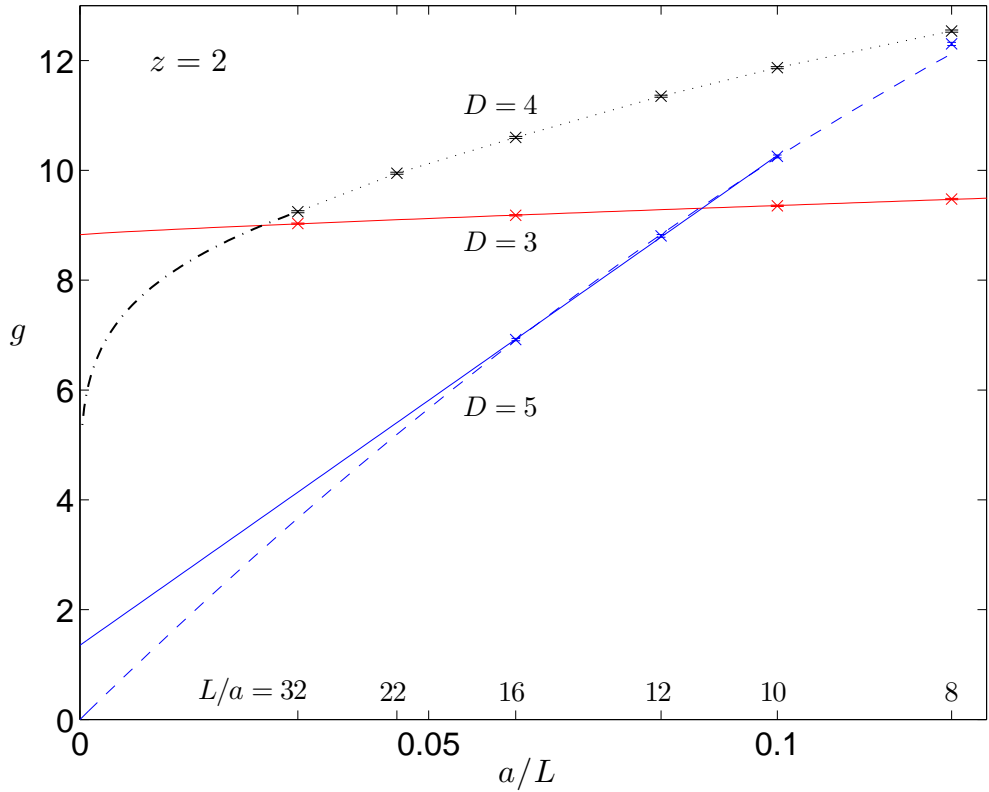}
\end{center}
  \caption{All measured couplings with fits (full and dashed lines) discussed
  in the text. Errorbars are barely visible within the symbols.\label{fig1}}
\end{figure}

In figure \ref{fig1} all coupling data are plotted. Fits (full and dashed
lines) all have acceptable $\chi^2$. The $D = 3$ values are almost cutoff
independent and approach a finite continuum value. The fit is $A + B (a /
L)^{\omega}$ with the corrections to scaling exponent
{\cite{Hasenbusch:1999mw}} $\omega = 0.85$. Due to the flatness, $\omega$ can
vary over a wide range including $\omega = 1$. For $D = 5$ the full line is $A
+ B a / L$ omitting the $L = 8$ lattice. The dashed fit is $A a / L + B (a /
L)^2$. A $32^5$ simulation would be of interest to better verify triviality
for this case. The $D = 4$ data show more curvature and the dotted lines just
connect the data points. A naive `eyeball' extrapolation in this plot against
$a / L$ would probably arrive at a nonzero value while on theoretical grounds
we shall argue now for the dash-dotted line extrapolating to the origin at a
vertical slope.

\begin{figure}[htb]
\begin{center}
  \includegraphics[width=0.8\columnwidth]{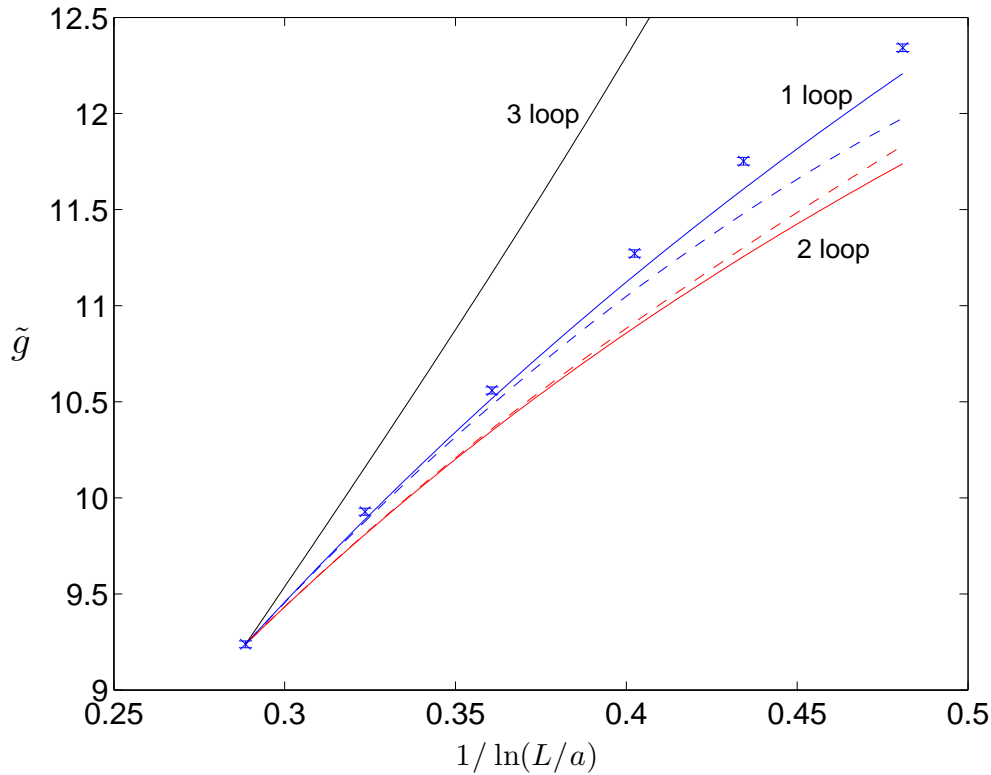}
\end{center}
  \caption{Numerical couplings together with the renormalization group
  evolution starting from the finest lattice. Dashed lines include lattice
  artifacts of the $\beta$-function.\label{fig2}}
\end{figure}

To confront the 4$D$ data with the theory {\cite{Brezin:1976bp}},
{\cite{Luscher:1987ay}} we plot against $[\ln (L / a)]^{- 1}$ in figure
\ref{fig2}. We solve the Callan Symanzik equation for the cutoff dependence of
the coupling starting from $L = 32.$ For this plot we have changed to the
coupling
\begin{equation}
  \tilde{g} = g [1 + (z a / L)^2 / 8]^{- 2} = g + \Omicron (a^2)
\end{equation}
differing by small cutoff effects only. For it in contrast to $g$ there are no
tree level artifacts in the $\beta$-function and also the cutoff corrections
of the one and two loop terms are more uniform. We have worked out the lattice
perturbation theory for our scheme up to two loops and could therefore, by
relating to {\cite{Luscher:1987ay}}, also obtain the three loop term (without
cutoff effects). Details will be reported elsewhere {\cite{PTUW}}. We here
draw the following conclusions: The one loop result is accurate to a few
percent for the scale changes considered. For instance, it accounts for 97\%
of the change $L / a = 32 \rightarrow 16$. The two loop term has a reasonable
relative size but the wrong sign. The three loop term is the first one that is
scheme dependent and hence depends on $z$. It is very large for $z = 2$ which
suggests that renormalized perturbation theory as an asymptotic expansion here
fails to improve the leading order. It rather is at its limit with only the
one loop approximation being numerically accurate at the percent level.
Nonetheless it seems convincing to now trust the one loop approximation for $a
< L / 32$ (for $z = 2$) which implies a vanishing $g$ in the continuum limit.
On the way to it, also the higher loop terms should eventually cooperate to
improve the approximation. The dash-dotted curve in figure \ref{fig1} shows
the one loop evolution continued. We plan a more detailed discussion of the
perturbative series and data for other $z$ values in {\cite{PTUW}}.

\begin{figure}[htb]
\begin{center}
  \includegraphics[width=0.8\columnwidth]{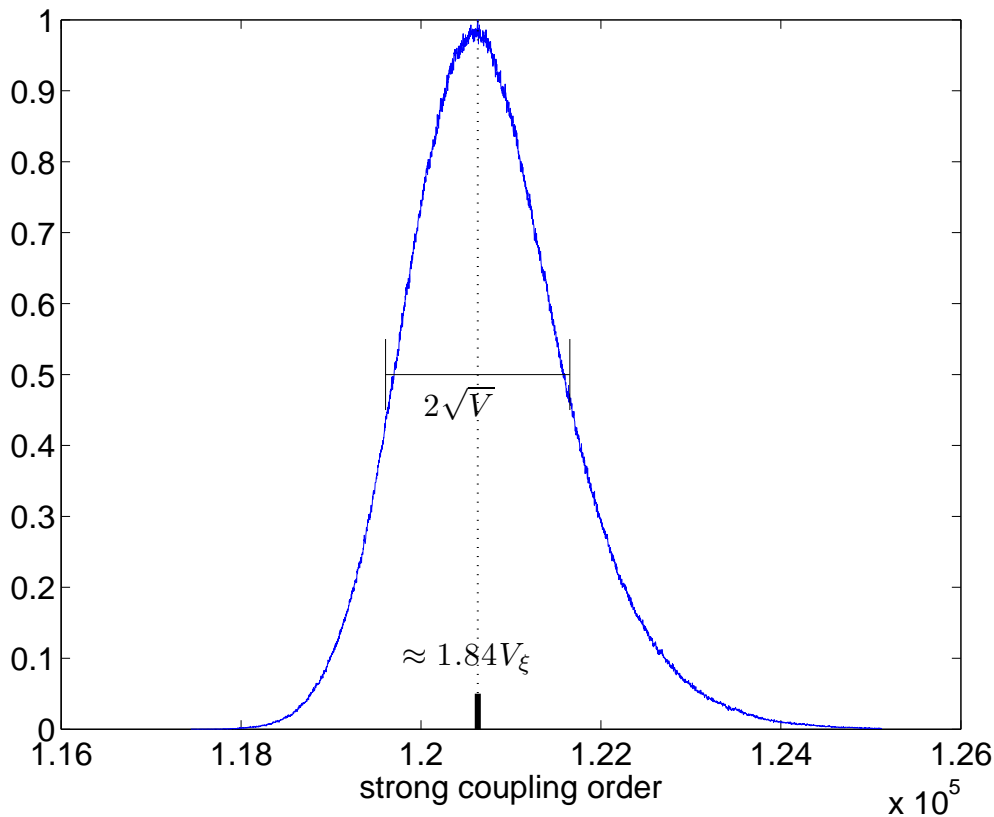}
  \caption{The order in $\beta$ of the diagrams sampled for $L = 32, D =
  4$ ($u=v$ only).\label{fig3}}
\end{center}
\end{figure}

We end on a more technical theme concerning the strong coupling simulation.
The order of the diagrams that the algorithm has picked to be important for
our physics is shown in figure \ref{fig3}. The peak is at the order of the
correlation volume $V_{\xi} = (a m)^{- 4}$ and the width seems to be
controlled by $(L / a)^2$.

{\noindent}\tmtextbf{Acknowledgments}. Exchange with Peter Weisz, Erhard
Seiler and Rainer Sommer was most beneficial to this project. Some part of the
effort was stimulated during a visit at the Max Planck (Werner Heisenberg)
Institut in M\"unchen by hospitality and support. I owe thanks to Willi Rath
for helping me with \tmtexttt{ranlux} under \tmtexttt{matlab} and to the HU physics
compute team for providing a smooth infrastructure. Finally financial support
of the DFG via SFB transregio 9 is acknowledged.

\end{document}